\documentclass[aps,showpacs,nofootinbib,superscriptaddress]{revtex4}
\usepackage{graphicx}

\begin{document}
\title{The nucleon axial mass and the MiniBooNE Quasielastic Neutrino--Nucleus Scattering problem}

\author{J. Nieves}
\affiliation{Instituto de F\'\i sica Corpuscular (IFIC), Centro Mixto
Universidad de Valencia-CSIC, Institutos de Investigaci\'on de
Paterna, E-46071 Valencia, Spain}
\author{I. \surname{Ruiz Simo}}
\affiliation{Departamento de F\'\i sica Te\'orica and IFIC, Centro Mixto
Universidad de Valencia-CSIC, Institutos de Investigaci\'on de
Paterna, E-46071 Valencia, Spain}
\author{M. J.  \surname{Vicente Vacas}}
\affiliation{Departamento de F\'\i sica Te\'orica and IFIC, Centro Mixto
Universidad de Valencia-CSIC, Institutos de Investigaci\'on de
Paterna, E-46071 Valencia, Spain}

\today

\begin{abstract}
The charged-current double differential neutrino cross section,
measured by the MiniBooNE Collaboration, has been analyzed using a
microscopical model that accounts for, among other nuclear effects,
long range nuclear (RPA) correlations and multinucleon scattering.  We
find that MiniBooNE data are fully compatible with the world average
of the nucleon axial mass in contrast with several previous analyses
which have suggested an anomalously large value.  We also discuss the
reliability of the algorithm used to estimate the neutrino energy.

\end{abstract}


\maketitle


Elastic neutrino nucleon scattering can be described by three dominant
form factors.  The two vector form factors $F_{1,2}(Q^2)$ are well
known from electron scattering (see, e.g. \cite{Perdrisat:2006hj}, for
a review).  The axial-vector form factor at $Q^2=0$, $F_A(0)$, is
determined from neutron $\beta$ decay. Assuming a dipole form, the
$Q^2$ dependence of $F_A = F_A(0)(1+Q^2/M_A^2)^{-2}$ can be
characterized by the axial mass $M_A$.  The value $M_A=1.03\pm 0.02$ GeV
is usually quoted as the world
average~\cite{Bernard:2001rs,Lyubushkin:2008pe}, although a recent
analysis claims an even smaller uncertainty ($M_A=1.014\pm 0.014$
\cite{Bodek:2007ym}).  It should be remarked that there are two
independent experimental sources of information for this parameter,
neutrino/antineutrino induced reactions and pion electroproduction. In
the first case, bubble chamber data for $\nu$-deuterium quasielastic
(QE) scattering play a dominant role.  The initial apparent
disagreement between the values of $M_A$ obtained with weak and
electromagnetic probes was solved after correcting for hadronic
effects~\cite{Bernard:2001rs} and now both sets of data are
consistent.  With these ingredients it looked straightforward to
describe $\nu$ QE scattering in nuclei with the high precision
required by the new and forthcoming neutrino experiments, that aim to
measure parameters such as the $\theta_{13}$ mixing angle or the
leptonic CP violation.

In this context, the charged current QE MiniBooNE
data~\cite{AguilarArevalo:2010zc} have been quite surprising.  First,
the absolute values of the cross section are too large as compared to
the consensus of theoretical models \cite{Boyd:2009zz,AlvarezRuso:2010ia}. Actually, the
cross section per nucleon on $^{12}C$ is clearly larger than for free
nucleons.  Second, their fit to the shape (excluding normalization) of
the $Q^2$ distribution leads to an axial mass, $M_A=1.35\pm 0.17$ GeV,
much larger than the previous world average.  In fact, the large value
of $M_A$ also implies a substantial increase in the total cross
section predicted by the Relativistic Fermi Gas model used in
the analysis, improving the agreement with the size of the cross
section.

Similar results have been later obtained analyzing MiniBooNE data with
more sophisticated treatments of the nuclear effects that work well in
the study of electron scattering. For instance,
Refs.~\cite{Benhar:2009wi,Benhar:2010nx} using the impulse
approximation with state of the art spectral functions for the
nucleons fail to reproduce data with standard values of $M_A$.  Large
axial mass values have also been obtained in
Ref.~\cite{Juszczak:2010ve} in a Fermi gas model and using spectral
functions and in Ref.~\cite{Butkevich:2010cr}, where data have been
analyzed in a relativistic distorted-wave impulse approximation and
with a relativistic Fermi gas model.

Certainly, there are some caveats that should be kept in mind like the
flux uncertainty or inadequacies on the subtraction of background
processes such as pion production. However, the associated
uncertainties have been estimated and  included in the
error bands provided in Ref.~\cite{AguilarArevalo:2010zc} and in the
previously quoted analyses.  Nonetheless, being the axial mass
relatively well established by electron data, the failure to describe
the MiniBooNE data with standard values of $M_A$ could point out more
to the incompleteness of the theoretical models than to the need of
reconsidering the value of the parameter.

As a consequence, several  approaches incorporating new mechanisms
that could contribute to the QE signal have been explored.  An
important step was undertaken in
Refs.~\cite{Martini:2009uj,Martini:2010ex} with the inclusion of two
nucleon mechanisms and other multinucleon excitations related to the
$\Delta$ resonance. These works could reproduce the MiniBooNE total QE
cross section without modifying the axial mass, suggesting that a good
part of the experimental cross section was not strictly QE
scattering. The importance of meson exchange currents and multinucleon
excitations has also been explored in Ref.~\cite{Amaro:2010sd}.  A
microscopic model for two nucleon excitation and pion production was
studied in Ref.~\cite{Nieves:2011pp}, supporting the findings of
Refs.~\cite{Martini:2009uj,Martini:2010ex}. This latter model was a
natural extension of the work in
Refs.~\cite{Nieves:2004wx,Nieves:2005rq,Gil:1997bm}, where the purely
quasielastic contribution to the inclusive electron and neutrino
scattering on nuclei had been analyzed.  The model includes one, two,
and even three-nucleon mechanisms, as well as the excitation of
$\Delta$ isobars.  There are no free parameters in the description of
nuclear effects, since they were fixed in previous studies of photon,
electron, and pion interactions with
nuclei~\cite{Gil:1997bm,Oset:1981ih,Salcedo:1987md,Carrasco:1989vq,Nieves:1991ye,Nieves:1993ev}.

In Refs.~\cite{Martini:2009uj,Martini:2010ex,Nieves:2011pp} only the
total cross section was evaluated and compared with the so called
``unfolded'' data of Ref.~\cite{AguilarArevalo:2010zc}.  Certainly,
the experimental data include energy and angle distributions and
therefore provide a much richer information. Furthermore, the unfolded
cross section is not a very clean observable after noticing the
importance of multinucleon mechanisms, because the unfolding itself is
model dependent and assumes that the events are purely QE.  The same
limitation occurs for the differential cross section $d\sigma/dQ^2$,
given that $Q^2$ is also deduced assuming the events are QE. From that
point of view, the best observable to compare with theoretical models,
and possibly constrain parameters, is the double differential cross
section $d^2\sigma/dT_\mu d\cos\theta_\mu$ because both the muon angle
and energy are directly measured quantities.  Our aim in this work, is
to analyse this latter observable within the theoretical model of
Refs.~\cite{Nieves:2004wx,Nieves:2005rq,Nieves:2011pp}. Full details
of the approach can be found there.

Here, we will briefly recall the main features of the model. It starts
from a relativistic local Fermi gas (LFG) picture of the nucleus,
which automatically accounts for Pauli blocking and Fermi motion. The
QE contribution was studied in Ref.~\cite{Nieves:2004wx} incorporating
several nuclear effects. The main one is the medium polarization
(RPA), including $\Delta$-hole degrees of freedom and explicit $\pi$
and $\rho$ meson exchanges in the vector-isovector channel of the
effective nucleon-nucleon interaction.  A correct energy balance is
imposed using the experimental Q values.  We will use here the full
relativistic model of Ref.~\cite{Nieves:2004wx} without the inclusion
of FSI interaction.  The reason is that FSI was implemented in a
nonrelativistic approach that makes it unsuitable for the large
momenta transferred that are reached in the experiment under study. As
it was discussed in ~\cite{Nieves:2004wx}, these FSI effects are
always smaller than 7 percent for the total cross
section\footnote{Thus, the results for the total cross section without
FSI are still inside the uncertainty band of
Ref.~\cite{Nieves:2011pp}} but could be more important in the angle
and energy distributions for the low neutrino energies studied in
~\cite{Nieves:2004wx}.  In Ref.~\cite{Benhar:2010nx}, it was found
that the main effect of FSI is a shift of $\sim 10$ MeV of the QE peak
for neutrino energies closer to the MiniBooNE neutrino flux mean
energy, $\langle E_\nu \rangle \sim 800 $ MeV, although that could
depend on details of the model~\cite{Amaro:2011qb}.

The model for multinucleon
mechanisms (not properly QE but included in the MiniBooNE
data~\cite{AguilarArevalo:2010zc}) has been fully discussed in
Ref.~\cite{Nieves:2011pp}. It contains some additional uncertainty
sources related to the detailed model for nucleon nucleon correlations, the
$\Delta$ axial couplings~\cite{Nieves:2011pp} or the use of inconsistent treatment of 
its vertices and propagator~\cite{Barbero:2008zza}.

To estimate the quality of the fit we use the following definition of
$\chi^2$ that properly takes into account the global normalization
uncertainty ($\Delta\lambda=0.107$) following the procedure of
~\cite{D'Agostini:1993uj},
\begin{equation}
 \chi^2=\sum_{i=1}^{137}
\left[\frac{\lambda\left( \frac{d^2\sigma^{exp}} {dT_\mu d\cos\theta }\right)_i  -  \left( \frac {d^2\sigma^{th} } {dT_\mu d\cos\theta } \right)_i }{ \lambda\Delta \left( \frac { d^2\sigma} { dT_\mu d\cos\theta} \right)_i } \right]^2 +
\left(\frac{\lambda -1}{ \Delta\lambda }\right)^2\,,
\end{equation}
where $\lambda$ is a global scale. $\frac{d^2\sigma^{exp}} {dT_\mu
d\cos\theta }$ is the experimental cross section and $ \Delta \left(
\frac { d^2\sigma} { dT_\mu d\cos\theta} \right)$ its uncertainty,
both taken from Ref.~\cite{AguilarArevalo:2010zc}.  The sum runs over
the 137 angle-energy bins with a cross section different from zero in
Ref.~\cite{AguilarArevalo:2010zc}.

As a first test, we have minimized $\chi^2$ as a function of the axial
mass in a simplified version of the model without multinucleon
mechanisms and without RPA. This model should be quite similar to the
one originally used in the MiniBooNE analysis. The main difference
being that we use a local rather than global Fermi gas in the
calculation.  We obtain $M_A= 1.32\pm 0.03$ GeV with $\chi^2=35$. The fit
is obviously very good and in agreement with
Ref.~\cite{AguilarArevalo:2010zc}. The fitted scale is
$\lambda=0.96\pm 0.03$ also supporting MiniBooNE findings, that a
shape-only fit was also consistent with the total cross section.

\begin{figure}[htb]
\includegraphics[width=0.95\textwidth]{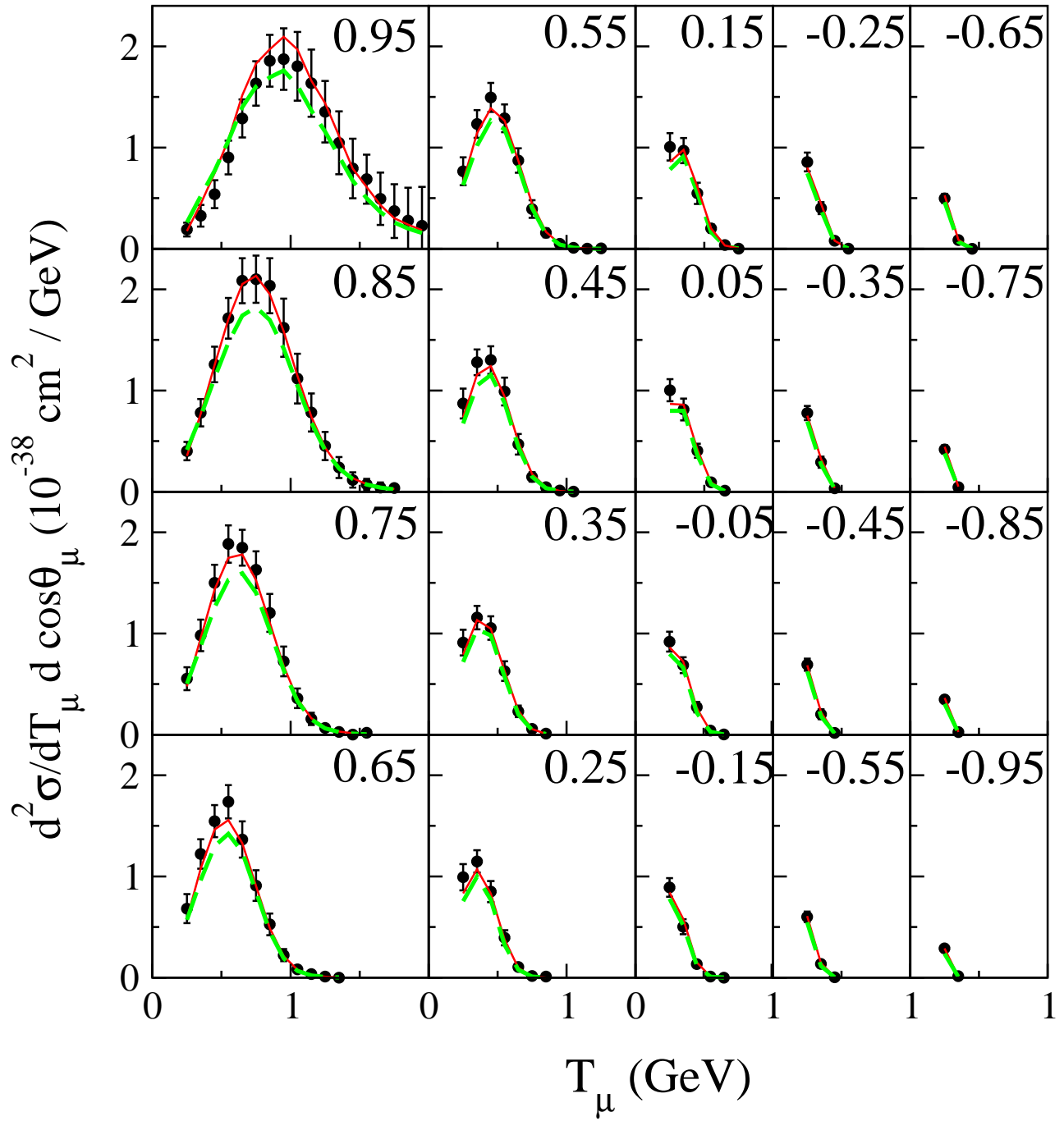}
\caption{Muon angle and energy distribution $d^2\sigma/d\cos\theta_\mu
 dT_\mu$.  Different panels correspond to the various angular bins
 labeled by their cosinus central value.  Experimental points from
 Ref.~\cite{AguilarArevalo:2010zc}. Green-dashed line (no fit) is the full
 model (including multinucleon mechanisms and RPA) and calculated with
 $M_A=1.049$ GeV.  Red-solid line is best fit ($M_A=1.32$ GeV) for the model
 without RPA and without multinucleon mechanisms.}
\label{fig:1}
\end{figure}

As a second test, we consider the full model with the same axial mass
used in our previous papers ($M_A=1.049$ GeV). The results corresponding
to these two versions of the model are shown in Fig.~\ref{fig:1}.  The
full model also agrees remarkably well with data. For this case we
have $\chi^2=52$ with $\lambda=0.89 \pm 0.01$.  This could look much
worse, but it is still a very good agreement with $\chi^2$ per degree
of freedom much lower than one and obtained without fitting any
parameter of the theoretical model.
Furthermore, the shape is very good and $\chi^2$ strongly depends on
the normalization (scale and axial mass are strongly correlated).
Therefore, from the quality of the fit only, one could not
discriminate between the two versions of the model. However, we should
recall that the RPA correlations and multinucleon mechanisms
correspond to real nuclear effects that must be incorporated in the
models.
\begin{figure}[htb]
\includegraphics[width=0.65\textwidth]{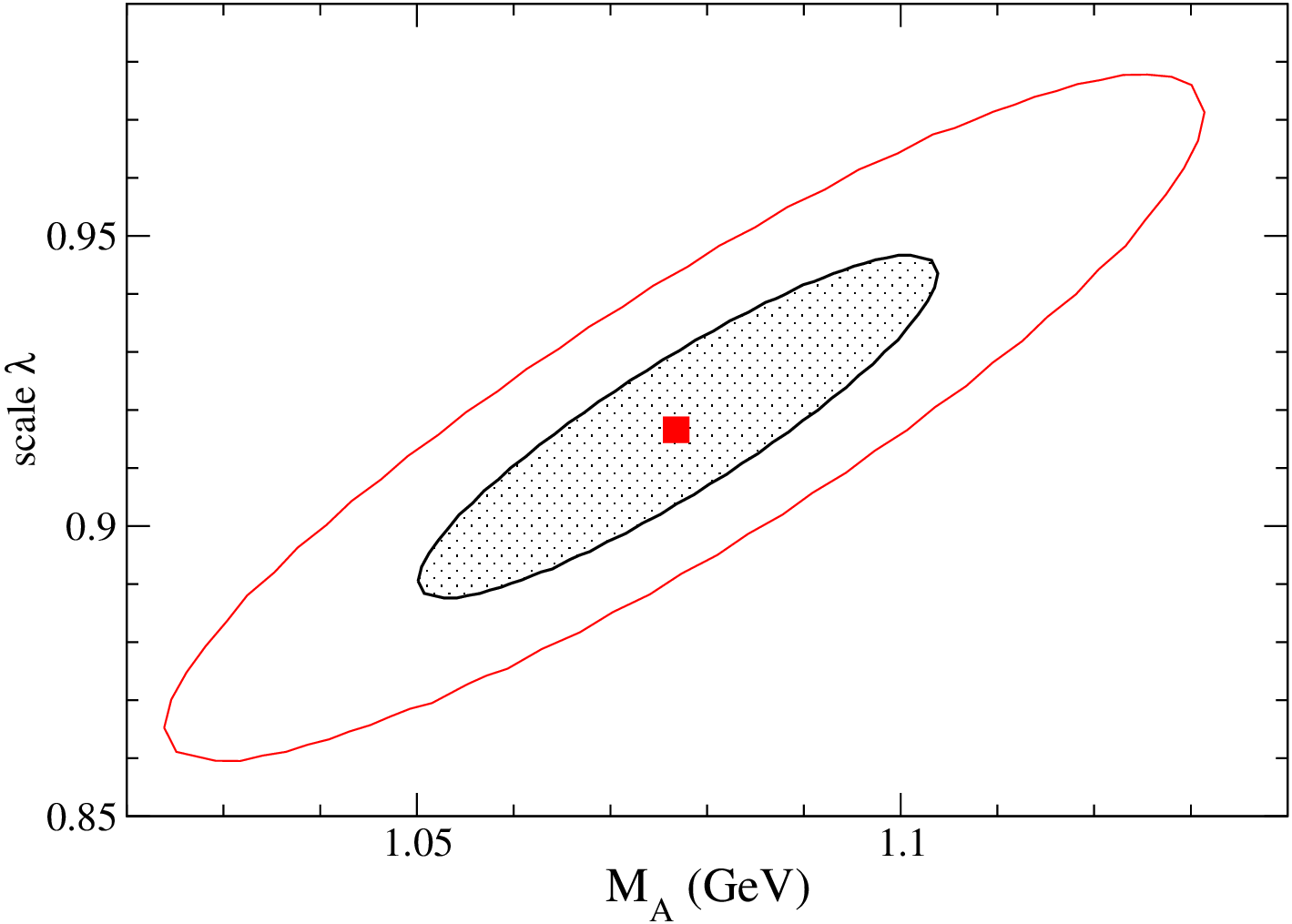}
\caption{Best fit value of $M_A$ and $\lambda$ and  1 and 2 $\sigma$ regions.}
\label{fig:2}
\end{figure}
Although, we think the consistency of MiniBooNE data with standard
values of $M_A$ has been established now, one could still go further and
 use our full model to fit the data letting $M_A$ to be a free
parameter.  We get $M_A=1.077\pm 0.027$ GeV and $\lambda=0.917\pm0.029$
with a strong correlation between both parameters. For this case,
$\chi^2=50$.  The 1 and 2 $\sigma$ contours are plotted in
Fig.~\ref{fig:2}. This is a somewhat large value for $M_A$ but we
think, the uncertainty size could be grossly underestimated. Notice
first that, in the absence of a proper correlation matrix, the
experimental uncertainties, except for the normalization, have been
treated as fully uncorrelated. In addition, one should include in the
minimization procedure not only the experimental but also the
theoretical uncertainties related to other parameters of the model
(e.g. $\pi NN$ form factors, short range correlations, $\Delta$ in
medium selfenergies, etc.).

\begin{table}[htb]
  \caption{Fit results for various models.  See description in the text.}
  \begin{tabular}{| l || r | r | r |}\hline
  Model & Scale &$M_A$ (GeV) & $\chi^2$/\# bins  \\ \hline
   LFG   & 0.96$\pm$0.03   & 1.321$\pm$0.030 & 33/137\\ \hline
  Full  & 0.92$\pm$0.03  & 1.077$\pm$0.027    &  50/137 \\ \hline
  Full, $q_{cut}=400$ MeV & 0.83$\pm$0.04  & 1.007$\pm$0.034    &  30/123\\ \hline
 \end{tabular}
\end{table}

The consideration of RPA and multinucleon mechanisms makes the present
model more appropriate than a pure impulse approximation for the low
momentum transfer region. Nonetheless, at very low momenta a more
detailed treatment of the nuclear degrees of freedom could be
necessary. As done in Ref.~\cite{Juszczak:2010ve}, we could exclude
from the analysis the bins with a large contribution of small momentum
transfer. There is some arbitrariness in the actual choice of the
cuts, but to allow for an easy comparison we have followed the
procedure of Ref.~\cite{Juszczak:2010ve} and implemented a transfer
momentum threshold $q_{cut}=400$ MeV. This eliminates 14 of the 137
measured bins (see Fig. ~3 from~\cite{Juszczak:2010ve}). The fitted
axial mass is then reduced to $M_A=1.007\pm 0.034$ GeV and
$\lambda=0.83\pm0.04$. As it is the case for the full calculation, the
inclusion of multinucleon mechanisms and RPA is essential to obtain
axial masses consistent with the world average.  For all cases the
best agreement is obtained for scale values lower than one. This is
even clearer for standard values of $M_A$. Whereas this possible
overestimation of the cross section could come from various sources
the simplest explanation is some underestimation of the neutrino flux.
\begin{figure}[htb]
\includegraphics[width=0.95\textwidth]{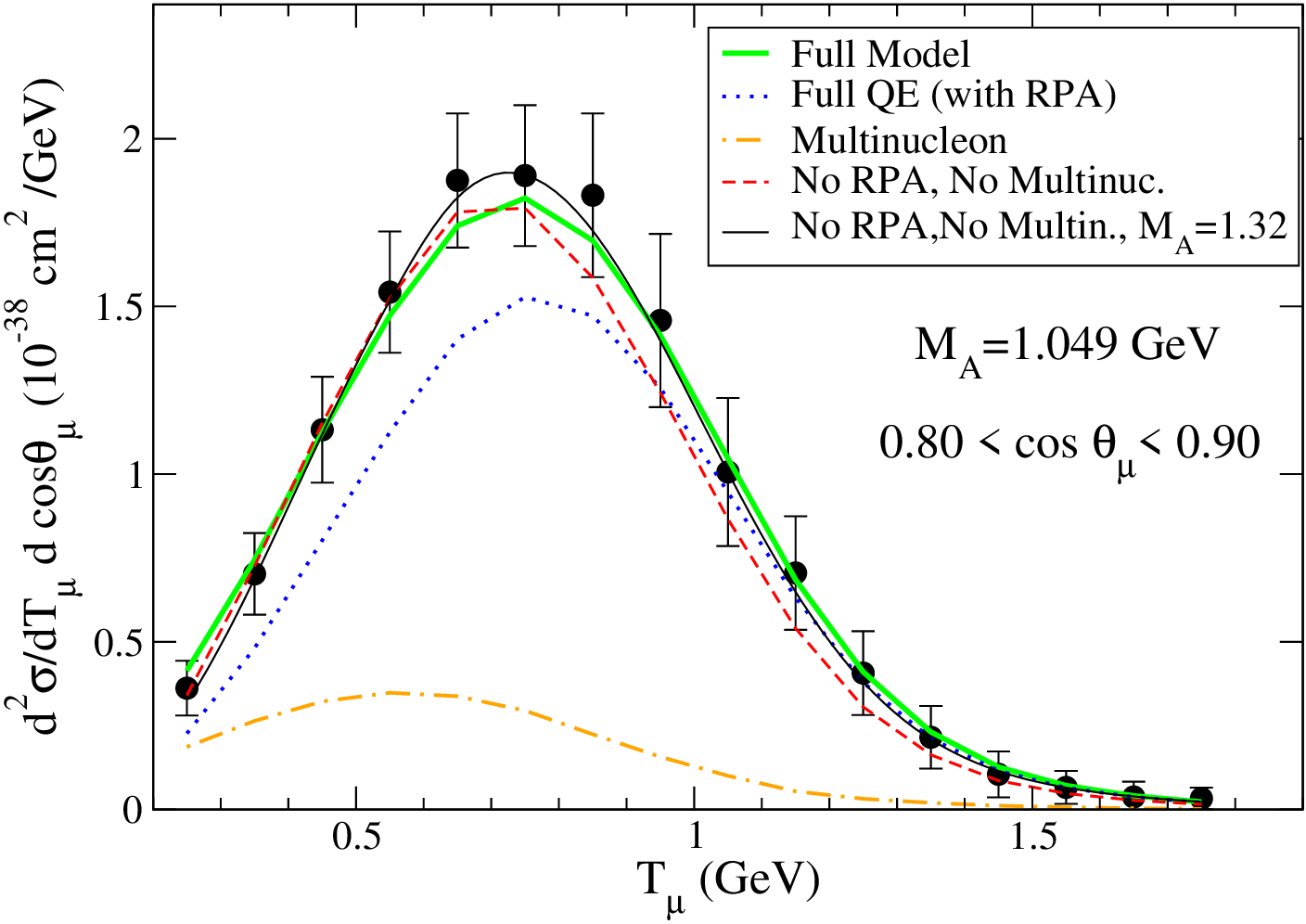}
\caption{Muon angle and energy distribution $d^2\sigma/d\cos\theta_\mu
dT_\mu$ for $0.80<\cos\theta_\mu<0.90$. Experimental data from
Ref.~\cite{AguilarArevalo:2010zc} and calculation with $M_A=1.32$ GeV
are multiplied by 0.9.  Axial mass for the other curves is $M_A=1.049$
GeV.}
\label{fig:3}
\end{figure}

Finally, in   Fig.~\ref{fig:3}, we show the contribution of the various mechanisms to the differential cross section at
$0.80<\cos\theta_\mu<0.90$. The experimental data have been scaled  to help in the discussion. The results of the LFG model (without RPA or multinucleon effects) with a large $M_A (=1.32)$ and with the same scale as data clearly provide an excellent fit, as it has been found by other groups.
For the rest of the curves we have taken $M_A=1.049$ GeV as in our previous
papers~\cite{Nieves:2004wx,Nieves:2005rq,Nieves:2011pp}. The LFG model with the low value of $M_A$, allowing for a 10 percent normalization uncertainty, also provides  an acceptable description of the data. One should remark that whereas this simple model agrees well for low and medium muon energies, it is systematically
below data at high energies.
The inclusion of collective effects
(RPA), dotted line, slightly improves the agreement at these high
energies. However, RPA strongly decreases the cross section at low
energies. Multinucleon mechanisms, which in average get a larger
energy transferred and thus accumulate their contribution at low muon
energies compensate that depletion.  Therefore, the final picture for
this observable is that of a delicate balance between a dominant
single nucleon scattering, corrected by collective effects, and other
mechanisms that involve directly two or more nucleons. As shown, both effects can be mimicked by using a large $M_A$ value.
 It is also
clear from this figure, that the proportion of multinucleon events
contributing to the ``QE'' signal is quite large for low muon energies
and thus, the algorithm commonly used to reconstruct the neutrino
energy is badly suited for this region. This could have serious
consequences in the determination of the oscillation parameters (see,
e.g., discussion in Ref.~\cite{Leitner:2010kp} and
Ref.~\cite{FernandezMartinez:2010dm}).

In summary, we have analyzed the MiniBooNE CCQE double differential
cross-section data using the theoretical model of
Refs.~\cite{Nieves:2004wx,Nieves:2005rq,Nieves:2011pp}. The model,
that starts from a relativistic local Fermi gas description of the nucleus,
includes RPA correlations and multinucleon effects. The same model is
quite successful in the analysis of nuclear reactions with electron,
photon and pion probes and contains no additional free parameters.
RPA and multinucleon knockout have been found to be essential for the
description of the data.  Our main conclusion is that MiniBooNE data
are fully compatible with former determinations of the nucleon axial
mass, both using neutrino and electron beams in contrast with several
previous analyses. The results also suggest that the neutrino flux
could have been underestimated.  Besides, we have found that the
procedure commonly used to reconstruct the neutrino energy for
quasielastic events from the muon angle and energy could be unreliable
for a wide region of the phase space, due to the large importance of
multinucleon events.

It is clear that experiments on neutrino  reactions on complex nuclei have reached a precision level that requires for a  quantitative description of sophisticated theoretical approaches. Apart from being important in the study of neutrino physics, these experiments are starting to provide very valuable information on the axial structure of hadrons.

\begin{acknowledgments}
 We thank L. Alvarez Ruso, R. Tayloe and G. Zeller for useful discussions.
This research was supported by DGI and FEDER funds, under contracts
  FIS2008-01143/FIS, FIS2006-03438, and the Spanish Consolider-Ingenio
  2010 Programme CPAN (CSD2007-00042), by Generalitat Valenciana
  contract PROMETEO/2009/0090 and by the EU HadronPhysics2 project,
  grant agreement n. 227431.  I.R.S. acknowledges support from the
  Ministerio de Educaci\'on.
\end{acknowledgments}


\end{document}